\documentclass[aps,prl,reprint,groupedaddress,showpacs]{revtex4-1}

\usepackage{graphicx}
\usepackage[ansinew]{inputenc}
\usepackage{array}
\usepackage{color}
\usepackage{amsmath}
\usepackage{amsxtra}
\usepackage{amstext}
\usepackage{amssymb}
\usepackage{latexsym}

\begin{document}


\title{Beyond the single-atom response in absorption lineshapes: Probing a dense, laser-dressed helium gas with attosecond pulse trains}

\author{Chen-Ting Liao}
\affiliation{College of Optical Sciences and Department of Physics, University of Arizona, Tucson, Arizona 85721, USA}

\author{Seth Camp}
\affiliation{Department of Physics and Astronomy, Louisiana State University, Baton Rouge, Louisiana 70803, USA}

\author{Kenneth J. Schafer}
\affiliation{Department of Physics and Astronomy, Louisiana State University, Baton Rouge, Louisiana 70803, USA}

\author{Mette B. Gaarde}
\email[]{gaarde@phys.lsu.edu}
\affiliation{Department of Physics and Astronomy, Louisiana State University, Baton Rouge, Louisiana 70803, USA}

\author{Arvinder Sandhu}
\email[]{sandhu@physics.arizona.edu}
\affiliation{College of Optical Sciences and Department of Physics, University of Arizona, Tucson, Arizona 85721, USA}

\date{\today}


\begin{abstract}

We investigate the absorption line shapes of laser-dressed atoms beyond the single-atom response, by using extreme ultraviolet (XUV) attosecond pulse trains to probe an optically thick helium target under the influence of a strong infrared (IR) field. 
{We study the interplay between the IR-induced  phase shift of the microscopic time-dependent dipole moment and the resonant-propagation-induced reshaping of the macroscopic XUV pulse. Our experimental and theoretical results show that as the optical depth increases, this interplay leads initially to a broadening of the IR-modified line shape, and subsequently to the appearance of new, narrow features in the absorption line.}

\end{abstract}



\maketitle



The development of attosecond ($10^{-18}$ sec) XUV spectroscopy\cite{Krausz2009} has opened new possibilities for application of transient absorption in real-time measurement and control of physical, chemical and biological processes at the natural timescale of electron motion \cite{Goulielmakis.2010.ATA.Kr.1st,Zenghu.2010.ATA.Ar}.
In particular, recent studies of attosecond transient absorption (ATA) in helium atoms have focused on how the presence of a moderately strong IR pulse alters the XUV absorption process at the single-atom level, for instance via the AC Stark-shifts \cite{Zenghu.2012.ATA.He.StarkShift}, or the appearance of new light-induced states \cite{LeoneKen.2012.ATA.He.LIS}. The ability of the IR field to alter the shape of the absorption profile from symmetric to dispersive or vice versa through a phase shift of the XUV-initiated time-dependent dipole moment has also been demonstrated \cite{Chu.2012.ATA.He,Pfeifer.2013.ATA.He.LorentzMeetsFano, Mette.Ken.2013.ATA.He.LIP, Pfeifer.2014.ATA.He, Delagnes2007}.

Most ATA experiments to date have assumed that the measured macroscopic optical density 
is directly proportional to the single-atom absorption cross section $\sigma(\omega)$ \cite{Leone.2008.ATA.Review, Zenghu.2012.ATA.He.StarkShift, Pfeifer.2013.ATA.He.LorentzMeetsFano, LeoneKen.2012.ATA.He.LIS, Keller.2013.ATA.He.ManyPlots}. This implicitly assumes Beer's law for dilute-gas absorption, $I_{out}(\omega) = I_{in}(\omega){\rm e}^{-\rho\sigma(\omega)z}$, where $\rho$ is the atomic density and $z$ is the propagation distance, even though many of these measurements were performed at relatively large optical densities, necessitated by the weak XUV sources. Only one experiment (with theory) has explicitly addressed propagation effects \cite{Leone.2013.He.ATA}, in addition to a number of calculations \cite{Mette.Ken.2013.ATA.He.LIP,Mette.Ken.2012.ATA.He.Reshaping, LinJphysB2012, Mette.Ken.2013.ATA.He.macroscopic3level}. {As attosecond science moves to more complex systems such as liquids, solids, and bio-materials which generally have higher densities \cite{Schultze.2013.ATA.SiO2, Jiang2014, Altucci2012}, it becomes imperative to understand macroscopic effects beyond Beer's law, and in particular how they can be disentangled from laser modification effects in ATA.}

\begin{figure}[t]
\includegraphics[width=0.49 \textwidth]{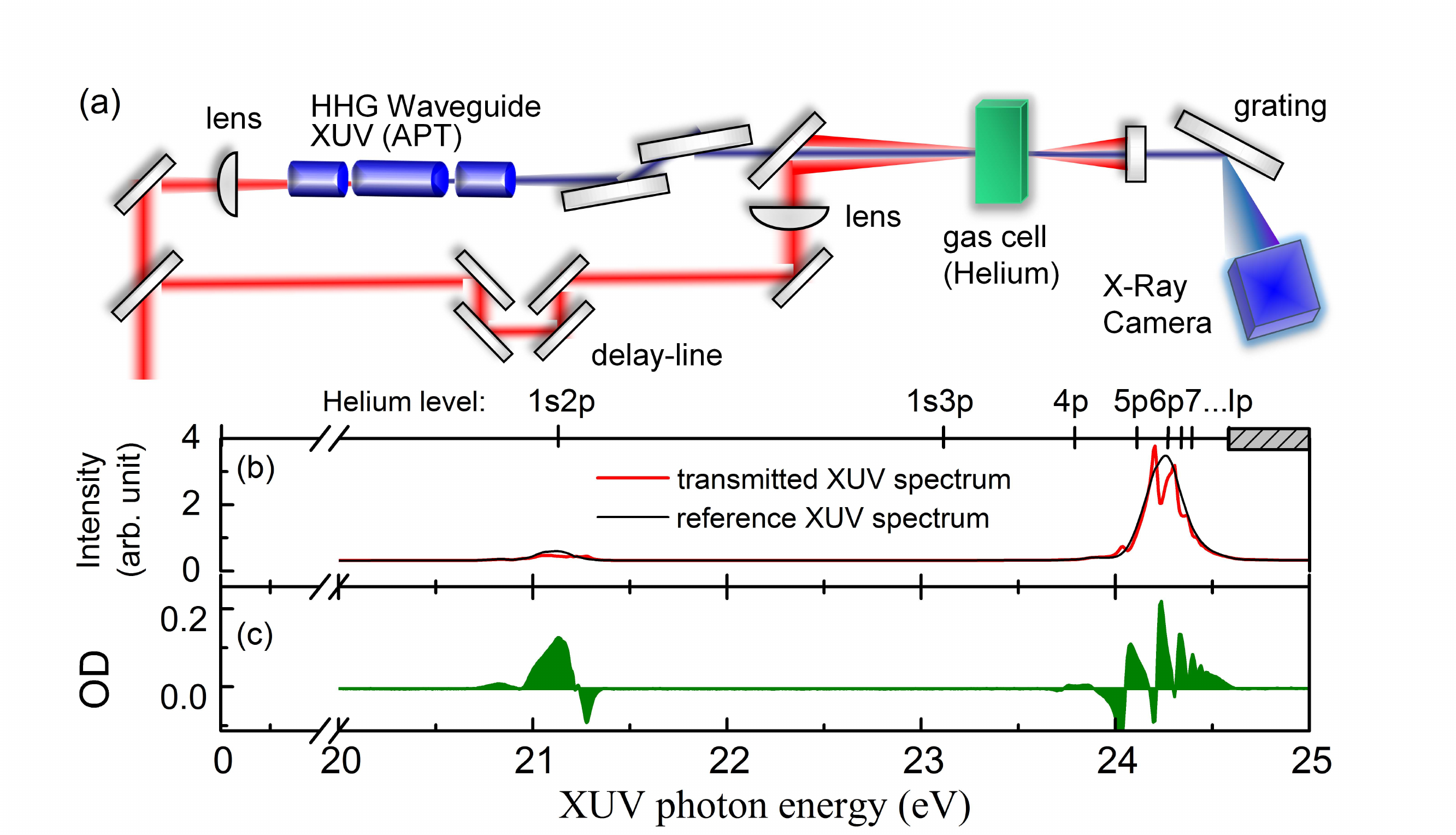} 
\caption{\label{fig:Fig1} 
(a) The experimental setup for ATA spectroscopy. (b) Typical XUV transmittance spectra with (red) and without (black) the IR field with Helium states labeled. (c) The optical density retrieved from (b).
}
\end{figure}

In this paper, we extend ATA spectroscopy to the more general case where collective effects enter into both experimental and theoretical considerations, by investigating the transient absorption of an attosecond pulse train (APT) in an optically thick helium sample interacting with a moderately strong IR pulse. {We consider the interplay between two different processes that can reshape the absorption line - the IR-induced phase shift of the time-dependent dipole moment \cite{Pfeifer.2013.ATA.He.LorentzMeetsFano, Mette.Ken.2013.ATA.He.LIP, Pfeifer.2014.ATA.He, Delagnes2007} and the macroscopic resonant propagation of the XUV pulse \cite{vanBurck.Review}. We show that initially, the two processes act independently so that the main effect of propagation is to broaden the characteristic line shape controlled by the IR perturbation (IRP)}. At higher pressures or longer propagation lengths, we observe new spectral features near the line center. {We establish the origin of these features in terms of the propagation-induced temporal reshaping of the short XUV pulse due to narrow-band resonances in He gas}. The temporal reshaping is a collective effect which has been observed and utilized in a range of applications from nuclear spectroscopy to quantum-well exciton studies and is often referred to as resonant pulse propagation (RPP) \cite{vanBurck.Review}. In this work, we explicitly observe its effects in ATA for the first time and show how the interplay between the single-atom IRP  and the collective RPP effects manifests itself in the absorption profile. 

Our experiment (Fig.\ref{fig:Fig1}(a)) utilizes Ti:Sapphire laser amplifier to produce 40 fs IR pulses at 1 kHz repetition rate with pulse energy 2 mJ, central wavelength 786 nm, and full width at half maximum (FWHM) bandwidth 26 nm. The XUV APT with $\sim$440 attosecond bursts and $\sim$4 fs pulse envelope is obtained by high harmonic generation in a Xenon filled waveguide. The APT is dominated by harmonics 13, 15, and 17. The XUV and IR pulses are combined using a mirror with a hole and focused into a 10 mm long He gas cell with a backing pressure of a few Torr, covered by Aluminum foil. The IR laser drills through the foil, allowing both XUV and IR beams to propagate. A spectrometer detects the transmitted XUV spectrum with a resolution of $\sim$7 meV at 24 eV.


Typical XUV transmission spectrum (Fig.\ref{fig:Fig1}(b)) measured at 8 torr backing pressure, with (black curve) or without (red curve) an IR pulse of 3 TW/cm$^2$ peak intensity. The XUV-IR delay is $\tau = +25$ fs, meaning XUV arrives after the IR. We average more than 10,000 laser shots with fluctuating carrier-envelope phase, so the observed absorption lineshapes represent an average over the sub-cycle delay variation.
As our spectrometer cannot resolve the narrow field-free absorption lines, the transmitted XUV spectrum $I_0$ in the absence of IR field is essentially the same as the input XUV spectrum. We therefore use it as a reference for evaluating the optical density, $OD(\omega,\tau) = -\log_{10}(I(\omega,\tau)/I_0)$, where $I(\omega,\tau)$ is the transmitted XUV spectrum in the presence of the IR \cite{Leone.2008.ATA.Review}. The spectral lineshape is strongly modified in presence of IR field, and both  absorption ($OD>0$) and emission ($OD<0$) features are observed (Fig.\ref{fig:Fig1}(c)). Moreover, the lineshapes are no longer simple Lorentzian-like peaks, but also include dispersive Fano-like profiles \cite{Fano1961}.

Our theoretical framework for ATA in laser-dressed He is described in \cite{Mette.Ken.2012.ATA.He.Reshaping}. Briefly, we numerically solve the coupled time-dependent Schr\"{o}dinger equation (TDSE), in the single active electron approximation, and the Maxwell wave equation (MWE). This yields the space- and time-dependence of {the full IR- and XUV-electric field} at the end of the gas. The optical density is then $OD(\omega) = -\log_{10} [I_{out}(\omega)/I_{in}(\omega)]$, where $I(\omega)$ is the radially integrated yield. We use an APT synthesized from harmonics 13 through 17, with initial relative strengths of 1:6:10 and all initially in phase. The FWHM duration of the APT is $\sim$5 fs, and the peak intensity is $10^{10}$~W/cm$^2$. The IR pulse has a central wavelength of 770 nm, a FWHM duration of 33 fs, and a peak intensity 2 TW/cm$^2$. For computational reasons, we consider a cell of length 1 mm and choose the range of pressures so that the evolution with pressure is similar to that of the experiment. A decay time of  $\sim$60 fs is imposed on the time-dependent polarization, to match the observed absorption line widths at low density, in the presence of the IR \cite{LeoneKen.2012.ATA.He.LIS}. 

\begin{figure}[t]
\includegraphics[width=0.49 \textwidth]{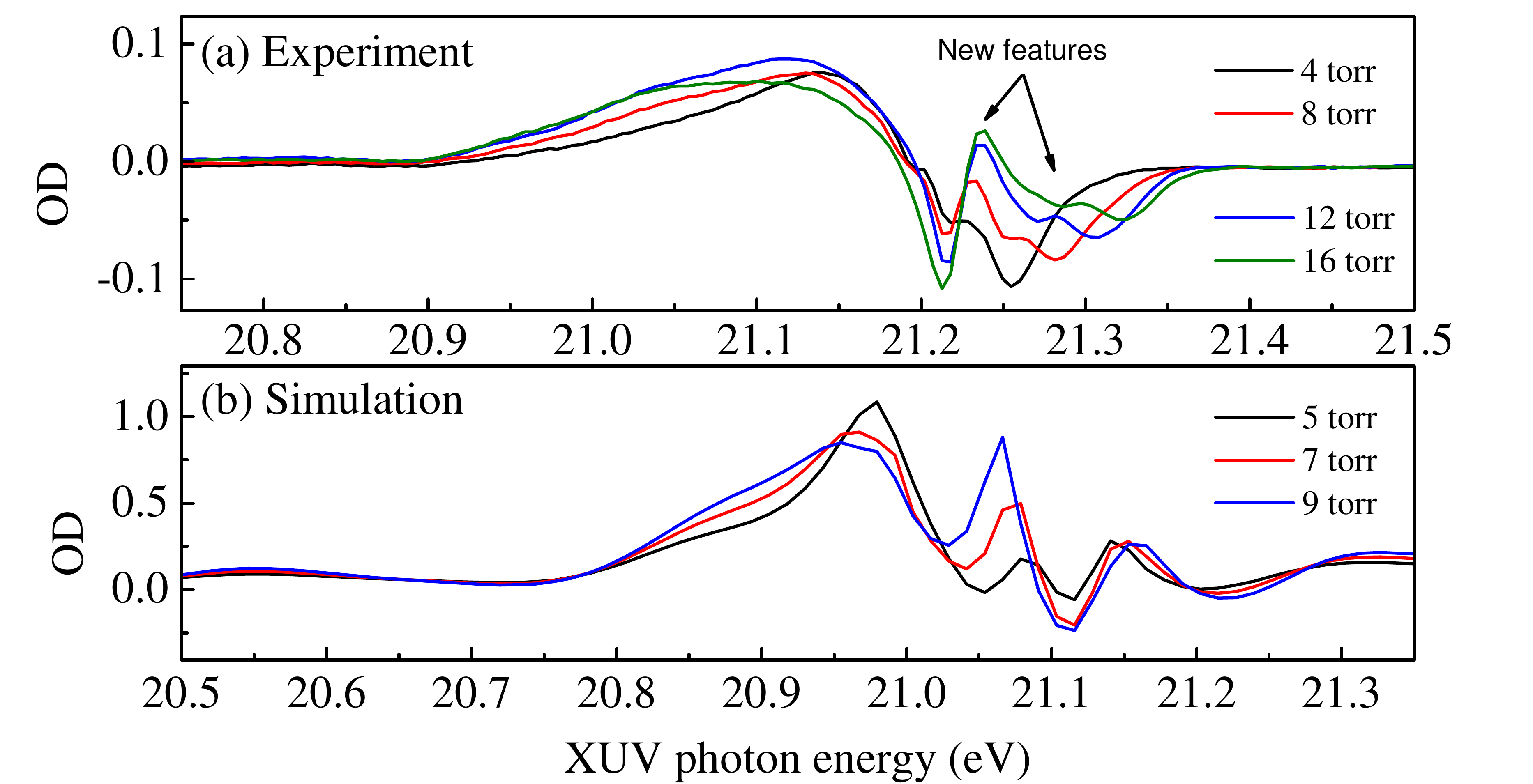} 
\caption{\label{fig:Fig2} 
Experimental (a) and theoretical (b) OD spectra around the He 1s2p state, in the presence of the IR field, and at different backing pressures. Arrows indicate new sharp spectral features.
}
\end{figure}

Fig.\ref{fig:Fig2}(a) shows the measured OD around the He 1s2p line for different backing pressures \cite{BackingPressure}, when the XUV-IR delay is approximately zero. Unlike ATA spectra in dilute gas (black curve) that exhibit a Fano-like profile with a broad peak and valley, we observe a complicated lineshape with finer absorption and emission features (colored curves). The lineshape strongly depends on the density of our thick sample, or more precisely, the pressure-length product in the gas cell. We observe broadening of the outer features of the lineshape, with the absorption peak on the left and the emission valley on the right both moving outward from the center as the pressure increases. In addition, we observe new sharp absorption features that appear on line center and move outward. 

Fig.\ref{fig:Fig2}(b) shows the OD calculated within the TDSE-MWE framework which in general agrees well with the experimental results {\cite{IRbackground}}. The energy axis of the theory curves has been shifted by about 0.1 eV to account for the small off-set of the 1s2p energy predicted by the pseudo-potential used in the TDSE calculations. To mimic the experimental delay resolution and carrier-envelope phase instability, the calculations have been averaged over a half IR cycle of delays around zero. 

\begin{figure}[t]
\includegraphics[width=0.49 \textwidth]{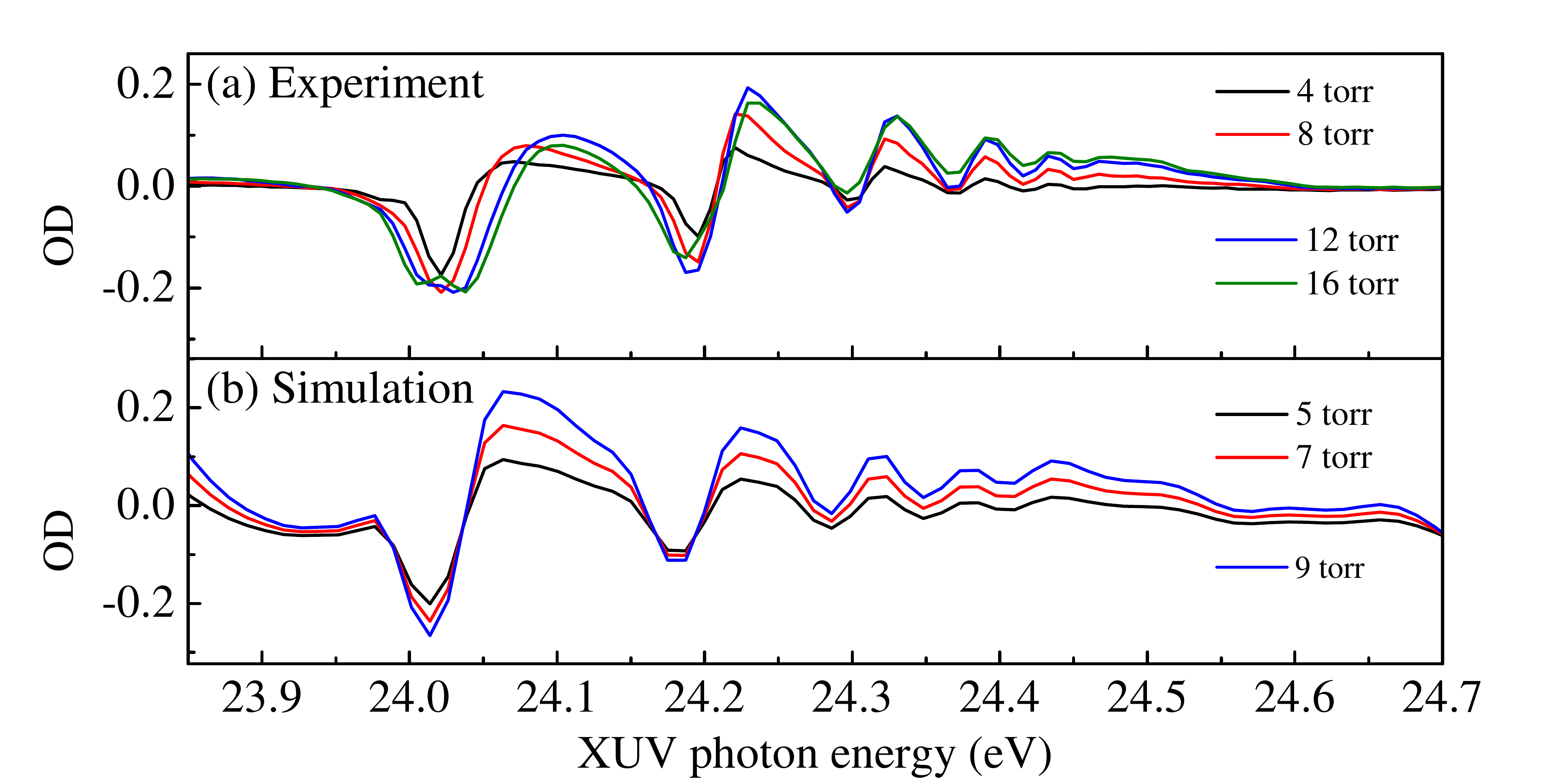} 
\caption{\label{fig:Fig3} Experimental (a) and theoretical (b) OD around the He 1snp states in the presence of IR.
}
\end{figure}
Fig.\ref{fig:Fig3}(a) and (b) show measured and calculated lineshapes of higher lying 1snp states,  where $n = 4,5,6,...$. Unlike the 1s2p state, these np lines show better defined Fano-like profiles, and the changes with pressure are less dramatic. Similar to the 1s2p line, the lineshapes are broadened, and both absorption and emission features are enhanced and move outward from the center of each resonance as the pressure increases. However, in contrast to the 1s2p line, we do not observe new sharp features.  

In the following, we will show that the spectral reshaping of the IR-controlled absorption profile originates in the RPP-induced temporal reshaping of the XUV pulse. To this end, we have performed a series of simpler TDSE-MWE calculations in which the TDSE is solved for a two-level system interacting with a resonant light field. We use an isolated 400 attosecond pulse centered on 21.1 eV, and we impose a $\sim$110 fs decay time on the time-dependent polarization.  The effect of the IR pulse is modeled as a time-dependent phase that accumulates on the upper state amplitude, proportional to the IR  intensity, as if the laser-imposed phase was strictly proportional to the Stark shift \cite{Mette.Ken.2013.ATA.He.LIP}. The IRP and the attosecond pulse begin at the same time. The IRP magnitude and the atomic density are chosen to match the experimentally observed line shape.

\begin{figure}[t]
\includegraphics[width=0.49 \textwidth]{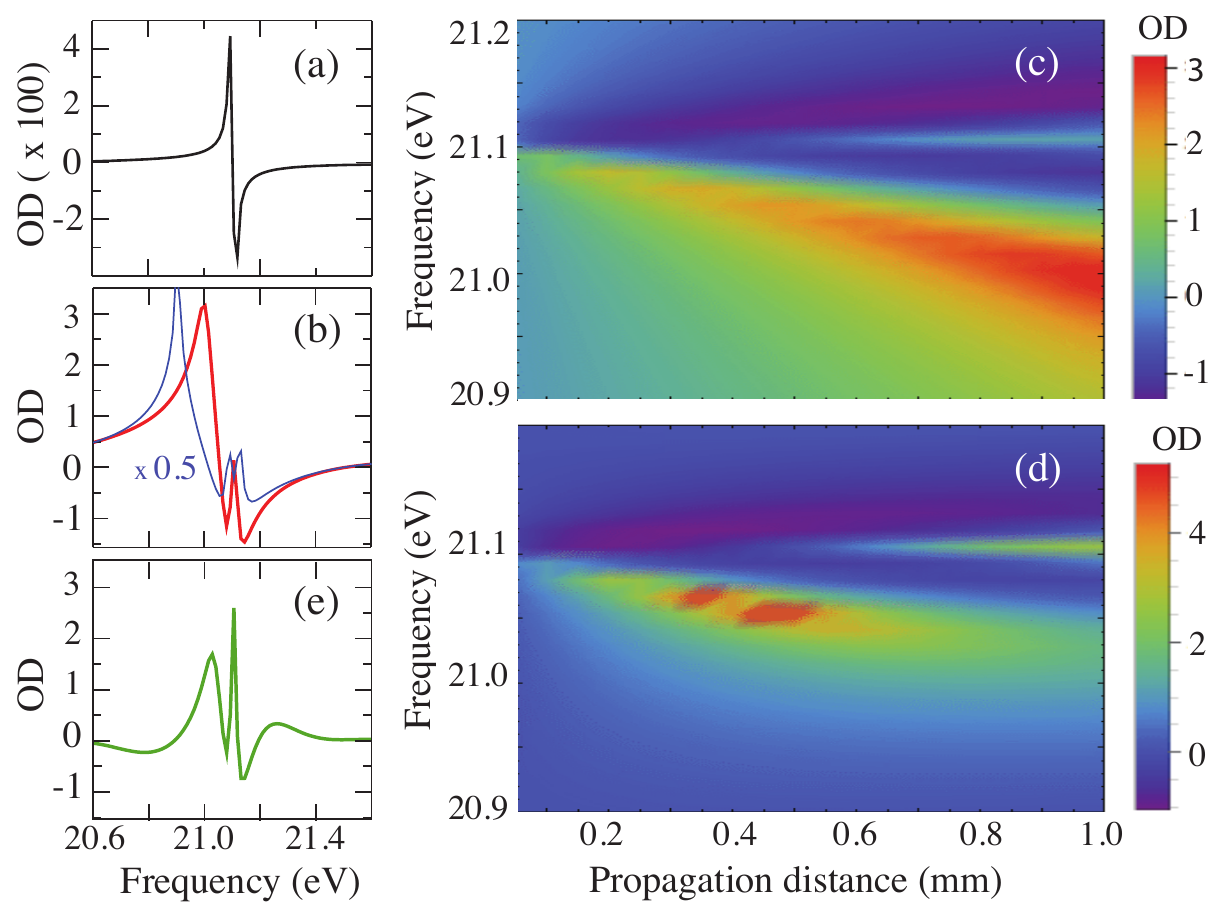} 
\caption{\label{fig:Fig4} 
Two-level TDSE-MWE calculations of OD. (a) and (b) show the initial (thick black line) and final (thick red line) OD for a 1 mm long medium with density $\rho=3\times10^{16}$ cm$^{-3}$, perturbed by a 2.7 fs IR pulse. Final OD for $\rho = 6\times10^{16}$ cm$^{-3}$  is also shown (thin blue line). (c) OD evolution with propagation distance, OD($z$), for $\rho=3\times10^{16}$ cm$^{-3}$ case in (a) and (b). (d) OD($z$) for $\rho=3\times10^{16}$ cm$^{-3}$ and a longer IRP (13 fs); the final OD for this is shown in (e).}
\end{figure}

Fig.\ref{fig:Fig4}(a-c) shows the evolution of the OD with propagation distance in a 1 mm medium with a density of $3 \times 10^{16}$ cm$^{-3}$, perturbed by a short, 2.7 fs, IR pulse. 
In the first half of the medium, the main effect of propagation is a linear broadening and increase of the initial dispersive-like line shape. In the second half of the medium, in addition to the broadening, we see the formation of a new sub-structure on line-center (21.1 eV). Further propagation (or higher density, thin blue line  in Fig.\ref{fig:Fig4}(b)) leads to additional broadening and the formation of a new substructure via splitting and separating of the first sub-structure into two. 

\begin{figure}[t]
\includegraphics[width=0.49 \textwidth]{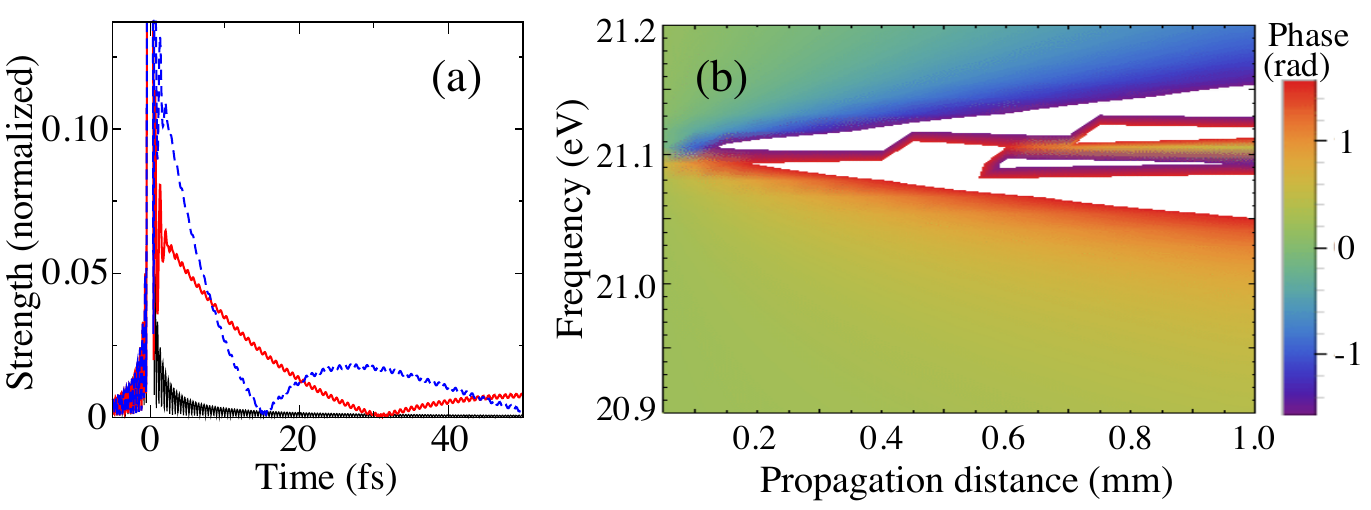} 
\caption{\label{fig:Fig5} Two-level TDSE-MWE results in the absence of IRP: (a) Normalized XUV electric field time structure at the beginning (thin black line) and end of the medium with $\rho=3\times10^{16}$ cm$^{-3}$ (thick red line), and $\rho=6\times10^{16}$ cm$^{-3}$ (dashed blue line). (b) The spectral phase accumulation vs. $z$. The white area marks where the spectral phase is larger than $\pi/2$ or smaller than $-\pi/2$.  
}
\end{figure}

Fig.\ref{fig:Fig5}(a) shows how the XUV pulse is temporally reshaped by RPP and develops a weak tail with longer and longer sub-pulses in it. As first discussed by Crisp \cite{Crisp1970.UltrafastPropagation}, the number of these sub-pulses increases and their duration decreases with propagation, as seen from the comparison of the low and high density cases. In the time domain, this RPP-induced reshaping can be understood as the consequence of the long tail of the electric field generated by the resonantly excited time-dependent dipole moment, the duration of which is determined by the effective lifetime of the upper level. This newly generated, long-lasting, electric field is out of phase with the driving, short-pulse, electric field, which leads to destructive interference (absorption) for a narrow range of frequencies. As the long tail of the electric field propagates through the medium, it in turn will excite new time-dependent dipole moments which will give rise to electric fields out of phase with that of the original tail (eventually giving rise to a new sub-pulse), and the process repeats. We note that this reshaping happens independently of the IR dressing field. The results in Fig.\ref{fig:Fig5}(a) were calculated in the absence of the IR, and are almost identical to the case when the IR is included. On the other hand, in the absence of reshaping, the IRP imposes a phase shift which builds up while the IR pulse is on, and then remains with the time-dependent dipole moment. The newly generated electric field will then no longer be exactly out of phase with the driving electric field. This manifests itself in the absorption spectrum as a more complex line shape in which some frequencies are absorbed and some are emitted, for instance as in Fig.\ref{fig:Fig4}(a), and where the line width is determined by the lifetime \cite{Pfeifer.2013.ATA.He.LorentzMeetsFano, Mette.Ken.2013.ATA.He.LIP, Pfeifer.2014.ATA.He}. 

Combining our understanding of the RPP and the IRP, we see that the primary effect of the RPP-induced reshaping is a shortening of the effective lifetime during which the IR-perturbed dipole moment oscillates, corresponding to the duration of the first sub-pulse in the XUV time profile \cite{vanBurck.Review}. This is because the phase of the electric field changes sign in the subsequent sub-pulse, thereby oscillating out of phase with the field that generated it which leads to renewed absorption and effectively ends the phase influence of the IRP. The RPP-shortened effective lifetime of the IRP causes the broadening of the IR-perturbed line shape. The new, narrow, absorption features are caused by the subsequent, longer, sub-pulse, and they do not (at first) alter the main IRP line shape. 

This explanation suggests that the IR-controlled change in the absorption profile remains a characteristic of the IRP as discussed in \cite{Pfeifer.2013.ATA.He.LorentzMeetsFano, Pfeifer.2014.ATA.He} as long as the effective, RPP-controlled, lifetime is much longer than the IR pulse. However, when the two time scales become comparable, this simple picture breaks down and the time-dependent dipole moment will exhibit a  more complicated amplitude and phase dependence. This is demonstrated in Fig.\ref{fig:Fig4}(d) which shows OD($z$) in the lower-density case where the IR pulse duration $\tau_{IR} = 13$ fs, which is close to the $1/e$ duration of the first sub-pulse, $T_{1/e}$, in the reshaped time-profile in Fig.\ref{fig:Fig5}(a) (thick red line). After $\approx 0.4$ mm of propagation (when $\tau_{IR} \approx 2T_{1/e}$), the linear increase of both the spectral width and the maximum and minimum values of the OD breaks down, and the narrow feature on line center strongly increases. This results in a final line shape (Fig.\ref{fig:Fig4}(e)) which is both narrower than what RPP-reshaping would predict, and very different from the initial line shape, with a strong central feature and several extra features in the wings.  
We observe similar effects for the 2.7~fs IR pulse for higher pressures, when the first sub-pulse becomes comparable to the IR pulse. In such cases, the relationship between the IRP and the absorption line shape is no longer simple, but depends explicitly on a self-consistent process in which the reshaped pulse and the IRP both play a role. 
In the experiment, this limit has been reached for the 1s2p absorption line shape at the highest pressures shown in Fig.~2, where the broadening saturates and the strength of the narrow, central, feature dominates.  
On the contrary, the line shape around the 1snp resonances in Fig.~3 is predominantly broadened because the single atom response is weaker for these transitions, and the temporal reshaping of the XUV pulse is less severe. 

Finally, we briefly present a frequency domain picture of the RPP-induced temporal reshaping. Fig.\ref{fig:Fig5}(b) shows the $z$-dependence of the XUV spectral phase accumulated due to dispersion. In the absence of the IRP, the spectral bandwidth over which significant phase is accumulated increases linearly. 
In the frequency domain, a phase change of more than $\pi$ will lead to a sign-change and thereby the formation of a new sub-pulse in the time domain. The temporal reshaping can thus equally well be thought of as being driven by dispersion in the frequency domain. In the absence of the IR, this broadening is primarily present in the spectral phase, but when the IR is present it also manifests itself as broadening of the absorption line width itself. A similar effect was observed in the strong-field limit in the absorption profile of a broadband pulse that was intense enough to induce Rabi cycling between the resonant levels \cite{Gaeta1998}.

The combined time- and frequency picture offers a guideline for when macroscopic effects cannot be ignored when interpreting an ATA line shape: when $T_{1/e}$ of the first sub-pulse becomes comparable to (approximately twice as long as) $\tau_{IR}$. The former can be estimated from the bandwidth $ \Delta\omega$ over which the accumulated spectral phase 
$\phi(\omega,z) =  \chi'(\omega) \omega/c z$ is larger (or smaller) than $\pm \pi/2$, as $T_{1/e} \approx \frac{2}{\pi \Delta\omega}$, where $\chi'(\omega)$ is the (IR- and spectrometer-broadened) susceptibility. {Under this criterion, the spectral phase on line center can still accumulate many factors of $\pi$. Therefore, this guideline on bandwidth} is much less restrictive on the experimental density than considering the first time the spectral phase $\chi'(\omega) \omega/c z$ equals $\pi$ which usually happens within tens of microns of propagation at typical ATA densities \cite{Leone.2013.He.ATA}.


In summary, we have studied ATA in laser-dressed He, in the limit where the gas medium is optically dense so that macroscopic effects influence the absorption line shapes. In both experiment and calculations we clearly observe the characteristic spectral signatures of temporal reshaping of the XUV pulse as it propagates through the resonant medium: the broadening of the bound state resonance profile as well as the emergence of new, narrow features at the resonance energy.
These results represent a novel manifestation of the general phenomenon of resonant pulse propagation \cite{vanBurck.Review}, in our case in the weak attosecond pulse regime, in which a moderately strong IR pulse facilitates the spectral measurement of an effect which is otherwise only apparent in the time-domain. 
The new regime of RPP can be applied for tailoring high frequency light fields\cite{Strasser2006}, thereby controlling light matter interaction on extremely fast timescales. Finally, we use our frequency-domain interpretation of this  phenomenon to propose a guideline for when macroscopic effects cannot be ignored in an ATA experiment: when the time scale of the first sub-pulse in the reshaped XUV field is comparable to the duration of the IR perturbation. This will be useful in a range of transient absorption experiments in which one aims to learn about the dynamics of a quantum system from the shape of the absorption profile, for instance under the influence of an IR pulse \cite{PfeiferNature2014}, but where experimental conditions necessitate using a relatively dense medium.

This work was supported by the National Science Foundation (NSF) under contract PHY-0955274, and by the U.S. Department of Energy, Office of Science, Office of  Basic Energy Sciences, under Award No. DE-FG02-13ER16403. High performance computational resources were provided by the Louisiana State University High Performance Computing Center.


\bibliography{RefsLiao3}

\begin{thebibliography}{27}
\bibitem [{Bac()}]{BackingPressure}%
  \BibitemOpen
  \href@noop {} {}\bibinfo {note} {The actual pressure in the interaction
  region is non-uniform, generally smaller than the backing pressure, and
  varies with the size of the laser-drilled entry and exit holes on the gas
  cell.}\BibitemShut {Stop}%
\bibitem [{IRb()}]{IRbackground}%
  \BibitemOpen
  \href@noop {} {}\bibinfo {note} {The absolute values of experimental and
  theoretical OD differ due to the IR background, which artifically reduces the
  experimental values obtained from the weak 13$^{th}$ harmonic.}\BibitemShut
  {Stop}%
\end{thebibliography}


\end{document}